\documentclass[iop]{emulateapj}
\usepackage{epstopdf}
\usepackage{subfigure}
\usepackage{amsmath}
\usepackage{color}

\usepackage{longtable}

\shorttitle{Non-thermal recombination}
\shortauthors{Reep \& Brown}

\begin{document}

\title{Amended results for hard X-ray emission by non-thermal thick target recombination in solar flares}

\author{J. W. Reep}
\affil{National Research Council Post-Doc Program, Naval Research Laboratory, Washington, DC 20375 USA}
\email{jeffrey.reep.ctr@nrl.navy.mil}
\and
\author{J. C. Brown}
\affil{Department of Physics and Astronomy, University of Glasgow, Glasgow G12 8QQ, UK}
\email{john.brown@glasgow.ac.uk}

\begin{abstract}
Brown \& Mallik 2008 and the Brown et al. 2010 corrigendum of it presented expressions for non-thermal recombination (NTR) in the collisionally thin- and thick-target regimes, claiming that the process could account for a substantial part of hard X-ray continuum in solar flares usually attributed entirely to thermal and non-thermal bremsstrahlung (NTB).  However, we have found the thick-target expression to become unphysical for low cut-offs in the injected electron energy spectrum.  We trace this to an error in the derivation, derive a corrected version which is real-valued and continuous for all photon energies and cut-offs, and show that, for thick targets, Brown et al. over-estimated NTR emission at small photon energies. The regime of small cut-offs and large spectral indices involve large (reducing) correction factors but in some other thick-target parameter regimes NTR/NTB can still be of order unity.  We comment on the importance of these results to flare and to microflare modeling and spectral fitting.  An empirical fit to our results shows that the peak NTR contribution comprises over half the hard X-ray signal if $\delta \gtrsim 6 (\frac{E_{0c}}{4\,\text{keV}})^{0.4}$.  
\end{abstract}

\keywords{atomic processes, Sun: corona, Sun: flares, Sun: X-rays, gamma rays}

\section{Introduction}
\label{sec:introduction}

Hard X-ray (HXR) emissions are widely used to understand the distribution and acceleration of electrons in solar flares.  For this reason, it is vitally important to understand correctly the radiative processes that contribute to the HXR bursts observed from flare electrons.  \citet{brown2008} (B08), \citet{brown2009} derived an equation for the emission of photons by two-body radiative recombination of non-thermal electrons onto hot ions in the thick-target approximation, which was updated in \citet{brown2010} (B10) to correct errors in the summation over ion types and in the atomic/ionic/radiative coefficients appropriate to the hydrogen ion species considered (Equation 3 of B10).  

By careful re-analysis of this work we found that their thick-target recombination expression gives imaginary, and thus unphysical, values for cut-off energies $E_{0c}<V_{Z_{\text{eff}}}/n^{2}$  (with $V_{Z_{\text{eff}}}$ the ionization energy of a bound state with effective charge $Z_{\text{eff}}$ and recombination end level quantum number $n$, in their notation), in  the intermediate range of this equation, $V_{Z_{\text{eff}}}/n^{2} < \epsilon < E_{0c} + V_{Z_{\text{eff}}}/n^{2}$ (for photon energy $\epsilon$).  This algebraic error in B08, carried forward into B10, would have been evident had B08 or B10 numerically plotted any case involving $E_{0c}<V_{Z_{\text{eff}}}/n^{2}$.

We therefore re-derive this Equation in Section \ref{sec:equations} to correct the functional form, in addition to the corrections already made by B10 to B08.  Spectra calculated with the amended equations are examined in Section \ref{sec:results}, to show that quite major corrections (reductions) in the predicted NTR occur for small cut-offs and large spectral indices, while for some other regimes NTR can be important.  We also discuss the errors that arise during spectral inversion when NTR is neglected in spectral fits, and important implications of NTR in Section \ref{sec:interpretations}.

\section{Equations}
\label{sec:equations}

We first consider recombination for one bound state of one ion, and then sum over ions $V_{Z_{\text{eff}}}$ and levels $n$ to obtain an expression for the total emission.  Let the following constants all be written together as $\Gamma_{Z\text{eff},n} = \frac{32 \pi r_{e}^{2} \chi^{2} Z_{\text{eff}}^{4}}{3 \sqrt{3} \alpha n^{3}}$, and then the cross-sectional area (cm$^{2}$) for recombination in the Kramers approximation \citep{kramers1923} is given by
\begin{equation}
Q_{R} = \frac{\Gamma_{Z\text{eff},n}}{\epsilon E}
\end{equation}
\noindent where $\epsilon$ is the energy of the emitted photon and $E$ the kinetic energy of the recombining electron.  We adopt the Kramers cross-section for simplicity of comparing NTR with NTB (and by analogy with B08),  although it is inappropriate for general use, when the Bethe-Heitler cross-section should be used \citep{bethe1934,koch1959}.  By Equation 7 of B08, the differential cross-section per unit $\epsilon$ is given by
\begin{equation}
\frac{d Q_{R}}{d\epsilon} = \Gamma_{Z\text{eff},n} \frac{\delta_{D}(E - \epsilon + V_{Z_{\text{eff}}}/n^{2})}{\epsilon E}
\end{equation}
\noindent where $\delta_{D}$ is the Dirac delta function and $V_{Z_{\text{eff}}}/n^{2}$ is the energy of the bound state of an ion with effective charge $Z_{\text{eff}}$ and quantum number $n$.  We then evaluate the photon yield $\eta_{RZ_{\text{eff}}}$ per unit $\epsilon$ for one electron of injection energy $E_{0}$ during its collisional lifetime using Equation B.7 of B08
\begin{align}
\eta_{RZ_{\text{eff}}}(\epsilon, E_{0}) &= \frac{1}{K} \int_{E} E \frac{dQ_{R}}{d\epsilon} dE  \nonumber \\
	&= \frac{\Gamma_{Z\text{eff},n}}{K} \int_{E} \frac{\delta_{D}(E - \epsilon + V_{Z_{\text{eff}}}/n^{2})}{\epsilon} dE  \nonumber \\
	&= \frac{\Gamma_{Z\text{eff},n}}{K} \times 
	\begin{cases}
	   0 & \text{if } E_{0} < \epsilon - \frac{V_{Z_{\text{eff}}}}{n^{2}} \\
	   1/\epsilon       & \text{if } E_{0} \geq \epsilon - \frac{V_{Z_{\text{eff}}}}{n^{2}}
	\end{cases}
\label{eta}
\end{align}
\noindent where $K$ is given by $2\pi e^{4} \Lambda$, $\Lambda$ being the Coulomb logarithm.  Note that the correct limits are in terms of $\epsilon - V_{Z_{\text{eff}}}/n^{2}$, not $\epsilon + V_{Z_{\text{eff}}}/n^{2}$ (cf. B08, B10).  We adopt a thick-target electron injection spectrum (electrons sec$^{-1}$ per unit $E_{0}$) of the same form as Equation B.10 of B08,
\begin{equation}
\mathfrak{F}_{0}(E_{0}) = (\delta - 1) \frac{\mathfrak{F}_{0c}}{E_{0c}} \times
	\begin{cases}
	\Big[\frac{E_{0}}{E_{0c}}\Big]^{-\delta} & \text{ if } E_{0} \geq E_{0c} \\
	0 & \text{ if } E_{0} < E_{0c}
	\end{cases}		
\label{electrondist}
\end{equation}
\noindent where $\delta$ is the spectral index, $\mathfrak{F}_{0c}$ is the electron rate above the cut-off (electrons sec$^{-1}$), and $E_{0c}$ is the low-energy cut-off.  We combine Equations \ref{eta} and \ref{electrondist} with Equation B.6 of B08 to determine the recombination spectrum $J_{RZ_{\text{eff}}}$ due to a given bound state of an ion.  However, the limits must be carefully accounted for.  We have two cases: $E_{0} < \epsilon - V_{Z_{\text{eff}}}/n^2$, in which case $J_{RZ} = 0$, and $E_{0} \geq \epsilon - V_{Z_{\text{eff}}}/n^{2}$ which must be split into three subcases: \textbf{1.} $E_{0} \geq \epsilon - V_{Z_{\text{eff}}}/n^{2} \geq E_{0c}$, \textbf{2.} $E_{0} \geq E_{0c} > \epsilon - V_{Z_{\text{eff}}}/n^{2} \geq 0$, and \textbf{3.} $\epsilon - V_{Z_{\text{eff}}}/n^{2} < 0$.

In the first case, $E_{0} \geq \epsilon - V_{Z_{\text{eff}}}/n^{2} \geq E_{0c}$, we have
\begin{align}
J_{RZ_{\text{eff}}}(\epsilon) &= \int_{E_{0} = \epsilon - V_{Z_{\text{eff}}}/n^{2}}^{\infty} \mathfrak{F}_{0}(E_{0})\ \eta(\epsilon, E_{0})\ dE_{0} \nonumber \\
	&= \frac{\Gamma_{Z\text{eff},n} (\delta - 1) \mathfrak{F}_{0c}}{K \epsilon E_{0c}} \int_{\epsilon - V_{Z_{\text{eff}}}/n^{2}}^{\infty} \Big[\frac{E_{0}}{E_{0c}}\Big]^{-\delta} dE_{0} \nonumber \\
	&= \frac{\Gamma_{Z\text{eff},n} \mathfrak{F}_{0c} }{K \epsilon} \Big[\frac{\epsilon - V_{Z_{\text{eff}}}/n^{2}}{E_{0c}}\Big]^{1-\delta}
\end{align}

In the second case, $E_{0} \geq E_{0c} > \epsilon - V_{Z_{\text{eff}}}/n^{2} \geq 0$, we similarly find
\begin{align}
J_{RZ_{\text{eff}}}(\epsilon) &= \int_{E_{0} = E_{0c}}^{\infty} \mathfrak{F}_{0}(E_{0})\ \eta(\epsilon, E_{0})\ dE_{0} \nonumber \\
	&= \frac{\Gamma_{Z\text{eff},n} \mathfrak{F}_{0c} }{K \epsilon}
\end{align}

In the third case, $\epsilon - V_{Z_{\text{eff}}}/n^{2} < 0$, there is no radiation since any photon emitted by recombination must have energy equal to $E + V_{Z_{\text{eff}}}/n^{2}$ for a given electron kinetic energy $E$, so $J_{RZ_{\text{eff}}} = 0$.  

Therefore, the combined expression is given by:
\begin{align}
J_{RZ_{\text{eff}}}(\epsilon) &= \frac{\Gamma_{Z\text{eff},n} \mathfrak{F}_{0c} }{K \epsilon}  \nonumber \\
	&\times\begin{cases}
	\Big[\frac{\epsilon - V_{Z_{\text{eff}}}/n^{2}}{E_{0c}}\Big]^{1-\delta} & \text{if } \epsilon \geq E_{0c} + \frac{V_{Z_{\text{eff}}}}{n^{2}} \\
	1 & \text{if } \frac{V_{Z_{\text{eff}}}}{n^{2}} \leq \epsilon < E_{0c} + \frac{V_{Z_{\text{eff}}}}{n^{2}} \\
	0 & \text{if } \epsilon < \frac{V_{Z_{\text{eff}}}}{n^{2}}
	\end{cases}
\end{align}
\noindent This function is real-valued for any positive values of $\epsilon$, $E_{0c}$, and $V_{Z_{\text{eff}}}/n^{2}$, and is continuous at the point $\epsilon = E_{0c} + V_{Z_{\text{eff}}}/n^{2}$.  

To calculate the entire spectrum due to multiple ions with many bound states, we have to sum over each ion and bound state.  The full spectrum is then given by (compare with Equation 3 of B10):
\begin{align}
J_{R}(\epsilon) &= \frac{32 \pi r_{e}^{2} \chi^{2} \mathfrak{F}_{0c} }{3 \sqrt{3} \alpha K \epsilon} \sum_{Z_{\text{eff}}} \sum_{n \geq n_{min}} p_{n} \zeta_{RZ_{\text{eff}}} \frac{1}{n^3}  \nonumber \\
	&\times\begin{cases}
	\Big[\frac{\epsilon - V_{Z_{\text{eff}}}/n^{2}}{E_{0c}}\Big]^{1-\delta} & \text{if } \epsilon \geq E_{0c} + \frac{V_{Z_{\text{eff}}}}{n^{2}} \\
	1 & \text{if } \frac{V_{Z_{\text{eff}}}}{n^{2}} \leq \epsilon < E_{0c} + \frac{V_{Z_{\text{eff}}}}{n^{2}} \\
	0 & \text{if } \epsilon < \frac{V_{Z_{\text{eff}}}}{n^{2}}
	\end{cases}
\end{align}
\noindent where $p_{n}$, $n_{min}$, and $\zeta_{RZ_{\text{eff}}}$ are defined in B10.  

In the range $V_{Z_{\text{eff}}}/n^{2} \leq \epsilon < E_{0c} + V_{Z_{\text{eff}}}/n^{2}$, the emission predicted by B10 is over-estimated by a constant factor of $\Big[\frac{E_{0c} - V_{Z_{\text{eff}}}/n^{2}}{E_{0c}}\Big]^{1-\delta}$.  For large cut-off energies and small spectral indices, this is relatively unimportant.  For large spectral indices, this term grows without bound and can lead to significant errors.  For small cut-off energies the discrepancy can be greater than 90\%, and the old expression gives complex, and thus unphysical, values for $E_{0c} < V_{Z_{\text{eff}}}/n^2$.  In Section \ref{sec:results} we briefly examine the effect this has on spectra.  

\section{Results}
\label{sec:results}

\begin{figure*}
\begin{minipage}[b]{0.5\linewidth}
\centering
\includegraphics[width=3.5in]{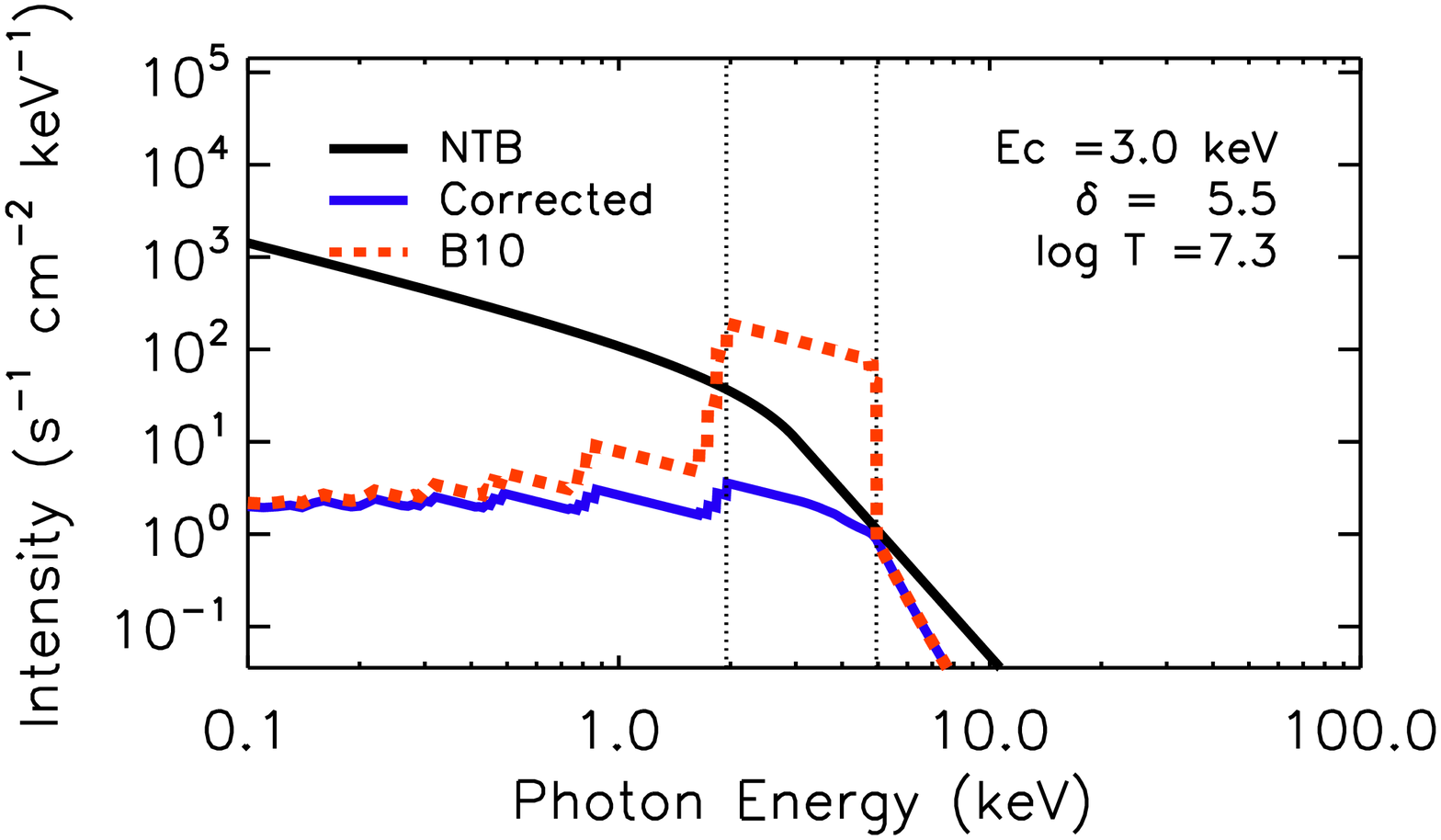}
\end{minipage}
\begin{minipage}[b]{0.5\linewidth}
\centering
\includegraphics[width=3.5in]{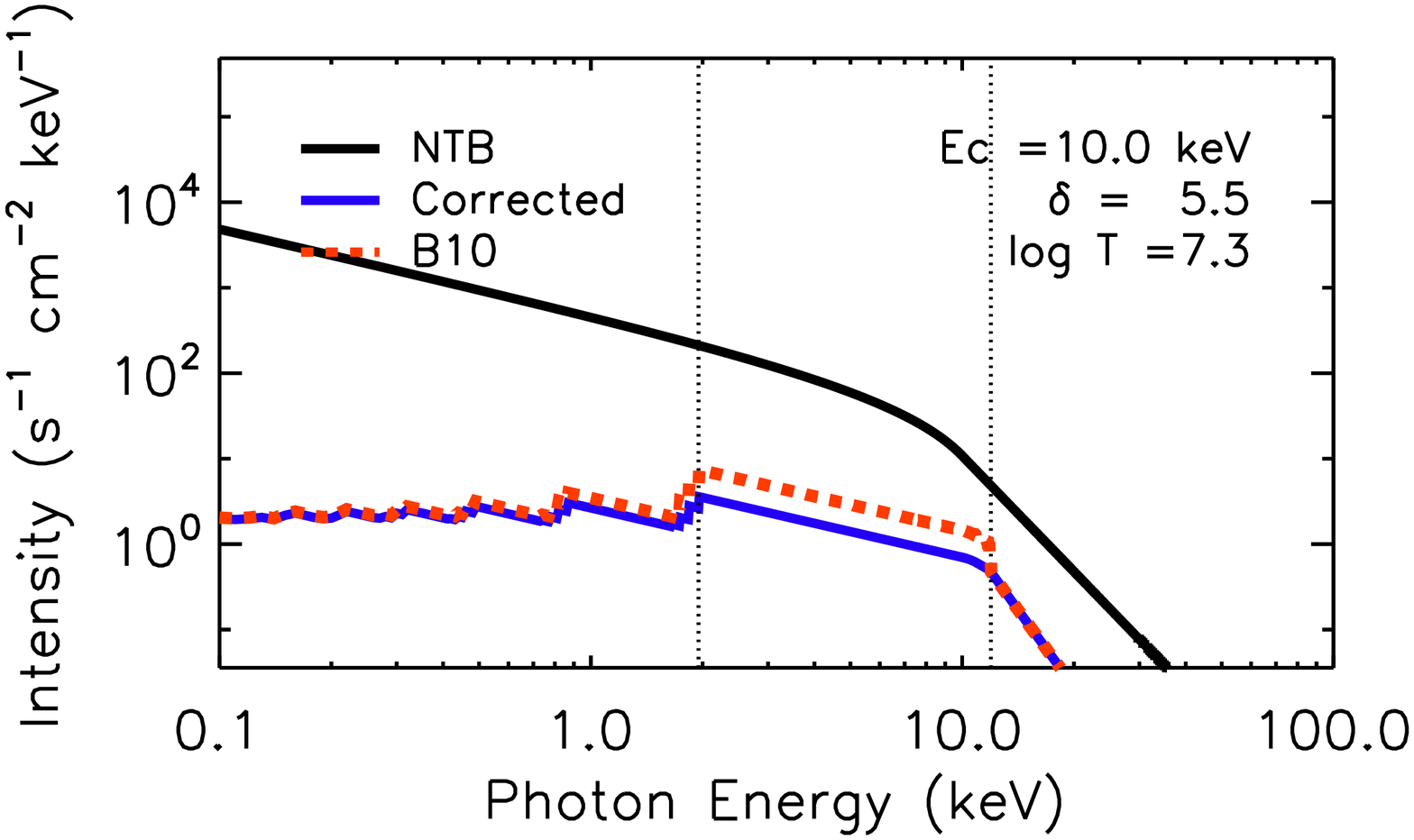}
\end{minipage}
\begin{minipage}[b]{0.5\linewidth}
\centering
\includegraphics[width=3.5in]{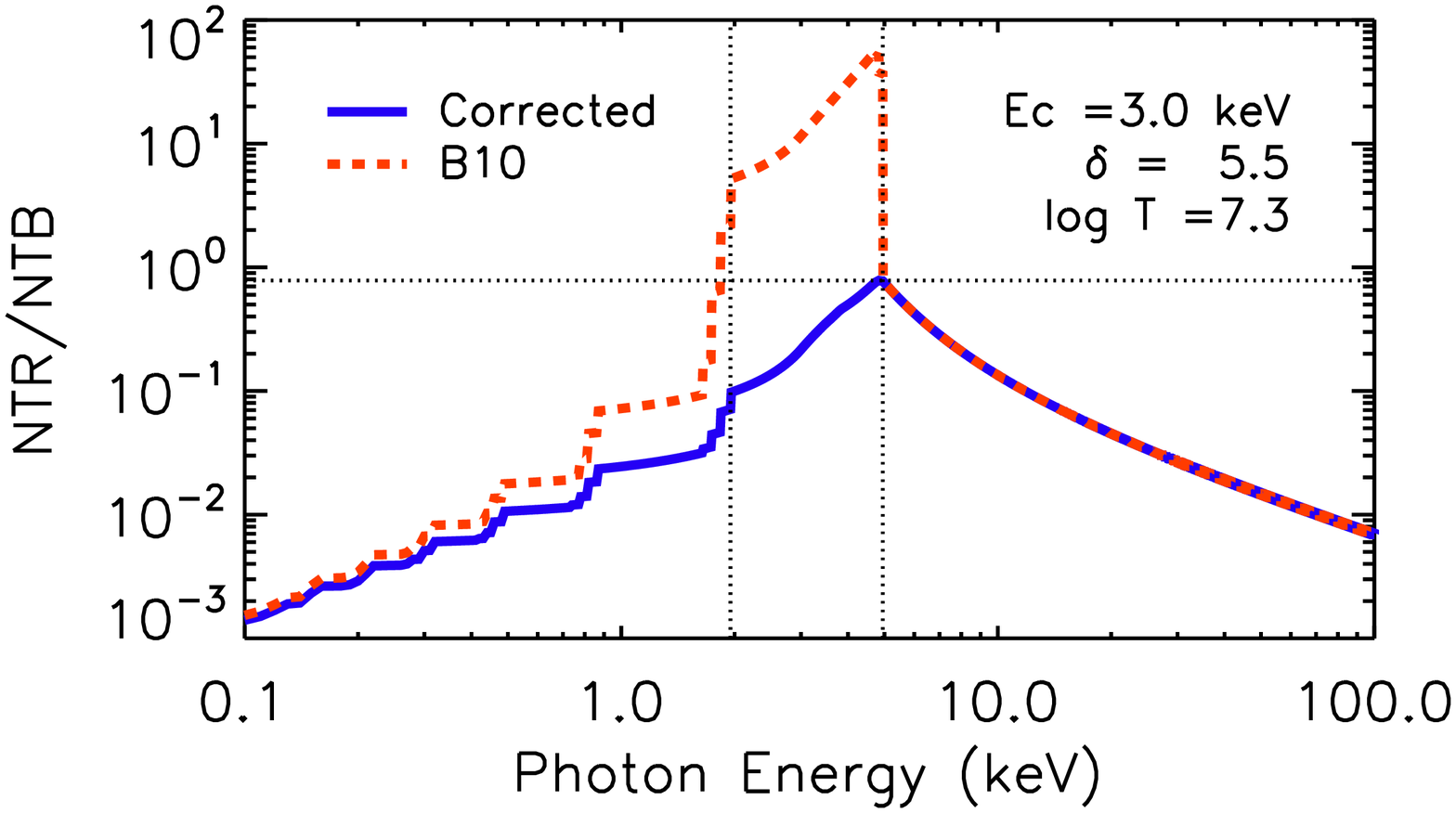}
\end{minipage}
\begin{minipage}[b]{0.5\linewidth}
\centering
\includegraphics[width=3.5in]{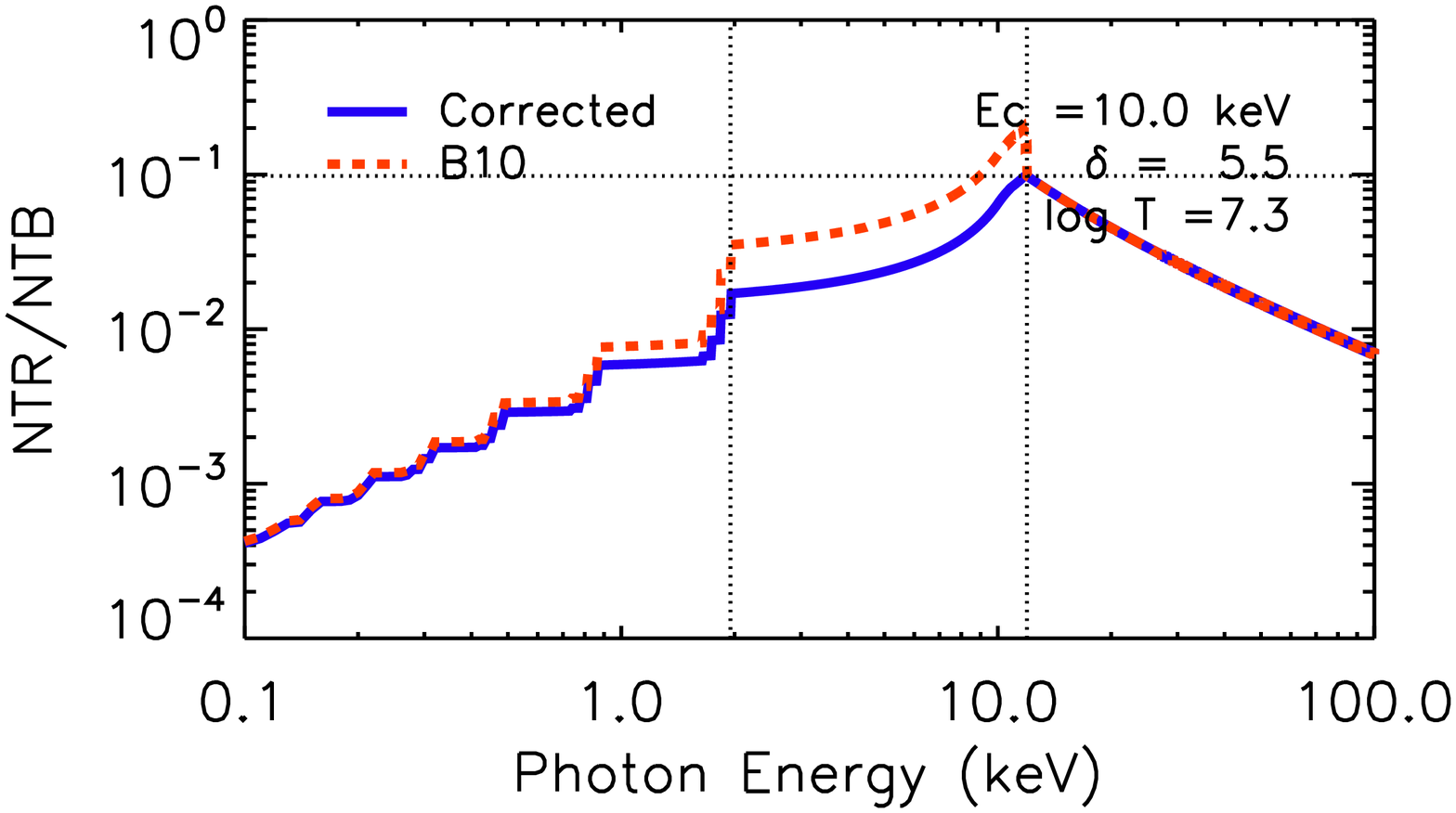}
\end{minipage}
\caption{Comparison of the corrected expressions for thick-target recombination to B10.  The top plots show the recombination spectra calculated with the amended and original expressions (along with NTB), and the bottom plots show the corrected and uncorrected ratios of non-thermal recombination to non-thermal bremsstrahlung.  The plots at left had a cut-off of 3\,keV, while the right had 10\,keV.  Both used a spectral index of 5.5, a temperature of 20\,MK, and an electron number flux above the cut-off of $10^{36}$ electrons sec$^{-1}$.  The vertical lines denote 1.95\,keV ($V_{Z_{\text{eff}}}/n_{min}^{2}$ for Fe XXV) and $E_{0c} + 1.95$\,keV, the range where the correction is most significant.  }
\label{comparison}
\end{figure*}

Using the (coronal) abundances and effective charges given in B08, along with ionization equilibria calculated with CHIANTI v.8 \citep{dere1997,delzanna2015}, we calculate example spectra to examine the amended equations.  Figure \ref{comparison} shows a comparison between the expressions of B10 and the amended ones here.  The top plots show the recombination spectra for cut-offs of 3 and 10\,keV, respectively, at a temperature of 20\,MK, spectral index of 5.5, and an electron rate above the cut-off $10^{36}$\,electrons sec$^{-1}$ (along with the NTB spectrum).  The bottom plots show the ratio of non-thermal recombination to non-thermal bremsstrahlung, calculated with equation B.11 of B08.  

The correction is most significant in the range $1.95$\,keV$< \epsilon \leq E_{0c} + 1.95$\,keV (1.95\,keV is $V_{Z_{\text{eff}}}/n_{min}^{2}$ for Fe XXV).  We have neglected Fe XXVI and XXVII, which should not be abundant at this temperature, so that above that range, the expressions are equivalent.  Below $1.95$\,keV, the correction becomes progressively smaller as photon energy decreases.  Comparing the two plots, it is clear that this correction is significant at low cut-offs, accounting for a factor of 100 error with a cut-off of 3 keV and only a factor of 2 with a cut-off of 10 keV.  
\begin{figure*}
\begin{minipage}[b]{0.5\linewidth}
\centering
\includegraphics[width=3.5in]{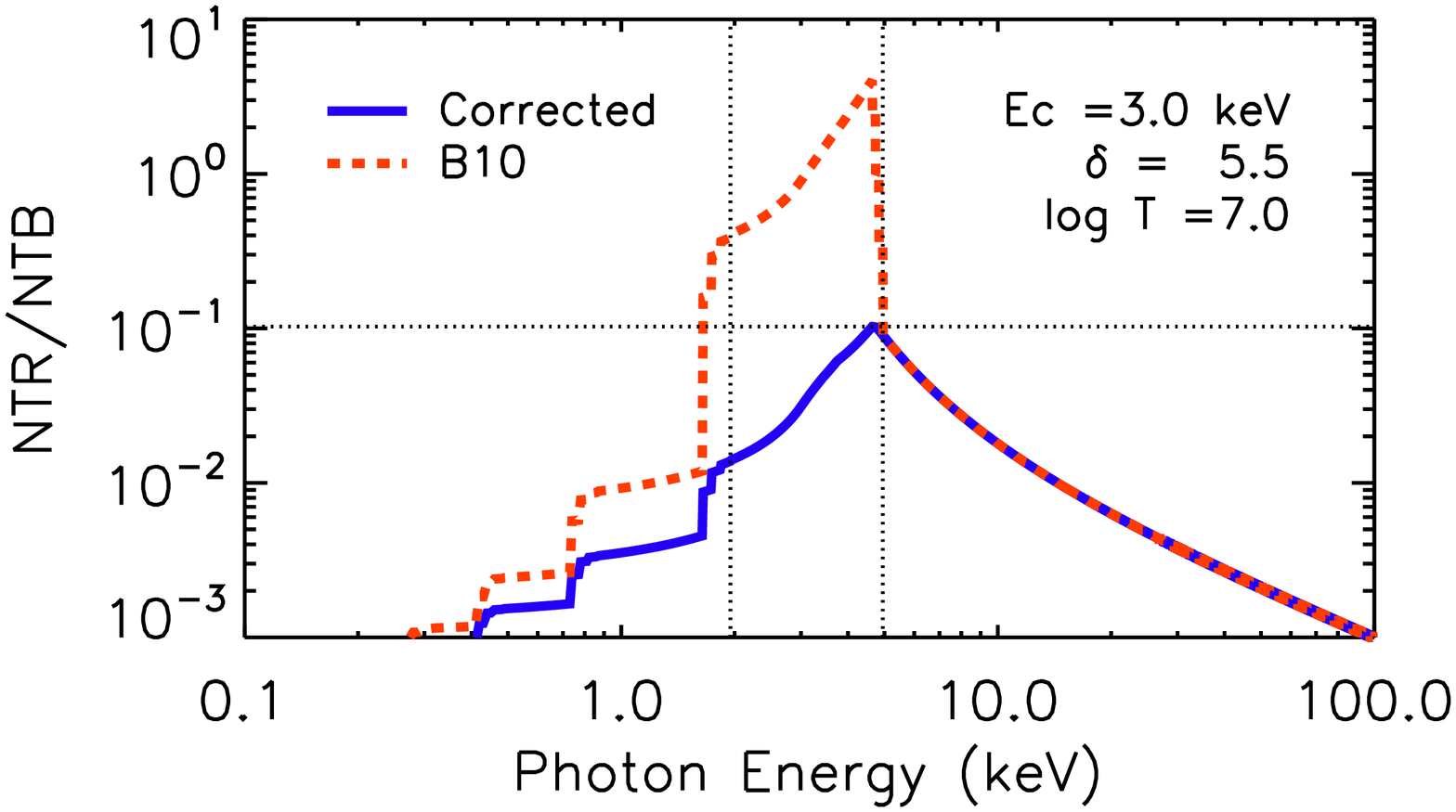}
\end{minipage}
\begin{minipage}[b]{0.5\linewidth}
\centering
\includegraphics[width=3.5in]{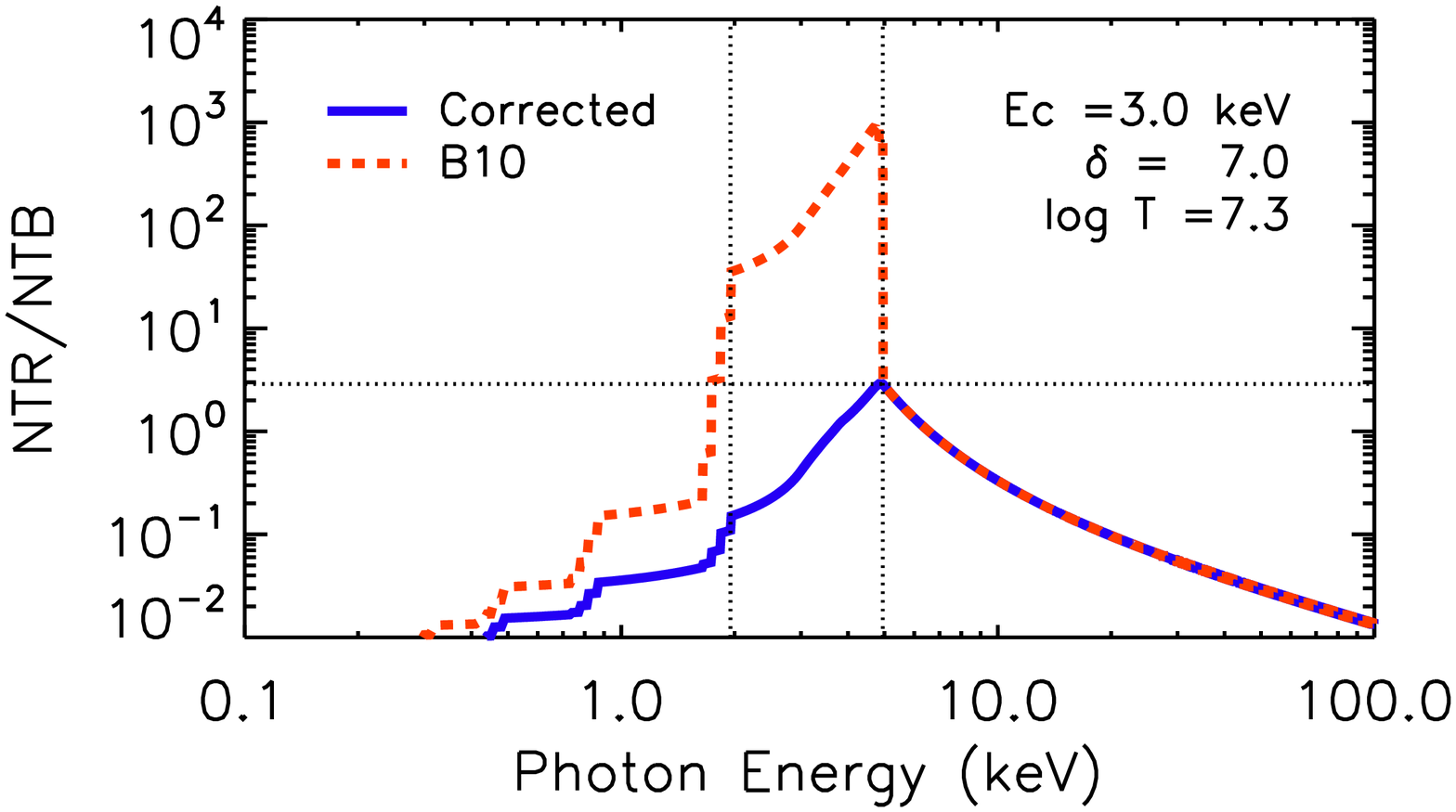}
\end{minipage}
\caption{The recombination/bremsstrahlung ratio for two more cases with (left) $\delta = 5.5$, $E_{0c} = 3$\,keV, and a temperature of 10 MK and (right) $\delta = 7$ and a temperature of 20 MK.  }
\label{othercases}
\end{figure*}

Figure \ref{othercases} shows two more examples, with a cut-off of 3\,keV once again.  The plot at left shows the recombination/bremsstrahlung ratio with a plasma temperature of 10\,MK (spectral index and electron flux the same as before), while the one at right shows the ratio for a spectral index of 7 (temperature 20\,MK).  It is immediately clear that the correction is more important at higher temperatures, where the hotter iron ions contribute more to the NTR spectrum.  On the other hand, for large spectral indices, the bremsstrahlung emissions diminish rapidly relative to recombination as photon energy decreases, so that recombination becomes increasingly dominant, and can be comparable to or greater than non-thermal bremsstrahlung.  

We have tabulated the peak of the ratio as a function of the spectral index and cut-off energy at 20 MK (Table \ref{ratiotable}).  We list the photon energy at which the ratio peaks (usually $E_{0c} + 1.95$\,keV) and the peak of the ratio itself.  At small cut-offs, it is clear that NTR is an important consideration regardless of the spectral index.  Similarly, at large spectral indices, it is clear that the process is important for all cut-offs.  An empirical fit shows that the peak NTR contribution comprises more than half of the HXR signal if $\delta \gtrsim 6 (\frac{E_{0c}}{4\,\text{keV}})^{0.4}$, or greater than 10\% if $\delta \gtrsim 4 (\frac{E_{0c}}{4\,\text{keV}})^{0.4}$.  Many flares follow a soft-hard-soft trend in spectral index ({\it e.g.} \citealt{grigis2004}), so that this process must be considered in spectral fitting, especially at early and late times.
\begin{deluxetable*}{c c c c c c c c c}
\caption{The NTR/NTB ratio as a function of $\delta$ (horizontal) and $E_{0c}$ (vertical) at a temperature of 20 MK.  Each grid cell shows the photon energy $\epsilon$ at which the ratio peaks, and the ratio itself at that energy.  If the ratio exceeds 10\%, it is shown in bold and blue to indicate where NTR becomes an important consideration. \label{ratiotable}} \\

\hline 
 & 3 & 4 & 5 & 6 & 7 & 8 & 9 & 10 \\
\hline
1 keV & 2.74, \color{blue} \textbf{0.16} & 2.83, \color{blue} \textbf{1.23} & 2.83, \color{blue}\textbf{6.66} & 2.83, \color{blue}\textbf{30.5} & 2.83, \color{blue}\textbf{127} & 2.83, \color{blue}\textbf{496} & 2.83, \color{blue}\textbf{1850} & 2.83, \color{blue}\textbf{6650} \\ 
2 keV & 3.74, 0.06 & 3.83, \color{blue}\textbf{0.31} & 3.83, \color{blue}\textbf{1.13} & 3.83, \color{blue}\textbf{3.48} & 3.83, \color{blue}\textbf{9.70} & 3.83, \color{blue}\textbf{25.4} & 3.83, \color{blue}\textbf{63.9} & 3.83, \color{blue}\textbf{156} \\ 
3 keV & 4.74, 0.04 & 4.83, \color{blue}\textbf{0.16} & 4.83, \color{blue}\textbf{0.48} & 4.83, \color{blue}\textbf{1.24} & 4.83, \color{blue}\textbf{2.88} & 4.95, \color{blue}\textbf{6.34} & 4.95, \color{blue}\textbf{13.4} & 4.95, \color{blue}\textbf{27.5} \\
5 keV & 6.74, 0.02 & 6.83, 0.07 & 6.83, \color{blue}\textbf{0.19} & 6.95, \color{blue}\textbf{0.42} & 6.95, \color{blue}\textbf{0.84} & 6.95, \color{blue}\textbf{1.56} & 6.95, \color{blue}\textbf{2.78} & 6.95, \color{blue}\textbf{4.80} \\
7 keV & 8.74, 0.01 & 8.83, 0.05 & 8.83, \color{blue}\textbf{0.11} & 8.95, \color{blue}\textbf{0.23} & 8.95, \color{blue}\textbf{0.43} & 8.95, \color{blue}\textbf{0.73} & 8.95, \color{blue}\textbf{1.21} & 8.95, \color{blue}\textbf{1.92} \\
10 keV & 11.83, 0.01 & 11.83, 0.03 & 11.95, 0.07 & 11.95, \color{blue}\textbf{0.13} & 11.95, \color{blue}\textbf{0.23} & 11.95, \color{blue}\textbf{0.37} & 11.95, \color{blue}\textbf{0.57} & 11.95, \color{blue}\textbf{0.86} \\
15 keV & 16.74, 0.01 & 16.83, 0.02 & 16.83, 0.04 & 16.95, 0.08 & 16.95, \color{blue}\textbf{0.12} & 16.95, \color{blue}\textbf{0.19} & 16.95, \color{blue}\textbf{0.28} & 16.95, \color{blue}\textbf{0.40} \\
20 keV & 21.74, $<$ 0.01 & 21.83, 0.01 & 21.95, 0.03 & 21.95, 0.05 & 21.95, 0.08 & 21.95, \color{blue}\textbf{0.13} & 21.95, \color{blue}\textbf{0.18} & 21.95, \color{blue}\textbf{0.25} \\
\hline
\end{deluxetable*}

To illustrate the importance of NTR in spectral fitting relative to NTB, we have synthesized photon spectra with NTB and NTR using the corrected formula and that of B10, in a case where NTR is expected to be important.  We adopt the following parameters: $E_{0c} = 3$\,keV, $\delta = 7$, an electron rate above the cut-off $\mathfrak{F}_{0c} = 10^{36}$\,electrons sec$^{-1}$, at a temperature of 20 MK.  We bin the energy using the commonly adopted RHESSI binning code 22, extended down to 1\,keV ({\it i.e.} 1/3 keV bins up to 15 keV, 1 keV bins up to 100 keV).  We then fit these photon spectra from 1-10\,keV with the IDL least squares fitting routine MPFIT \citep{markwardt2009}, assuming only contributions from NTB.  In this way, we gain a sense of the importance of ignoring NTR in spectral fits.  

Although thermal emissions generally dominate flare spectra at these energies, we are primarily concerned with the relative importance of NTR compared with NTB.  It is important to note that thermal emissions are weak early in the impulsive phase in some small events due to a small emission measure ({\it e.g.} \citealt{hannah2008,oflannagain2013}), and that they are weak in so-called cold flares ({\it e.g.} \citealt{fleishman2011,masuda2013,fleishman2016}), so that there are times when thermal emissions are relatively less important compared to non-thermal processes.  Further, while the global emission measure may be large, that does not preclude the possibility of emission from part of the volume being dominated by non-thermal emissions.

\begin{figure}
\centering
\includegraphics[width=3.5in]{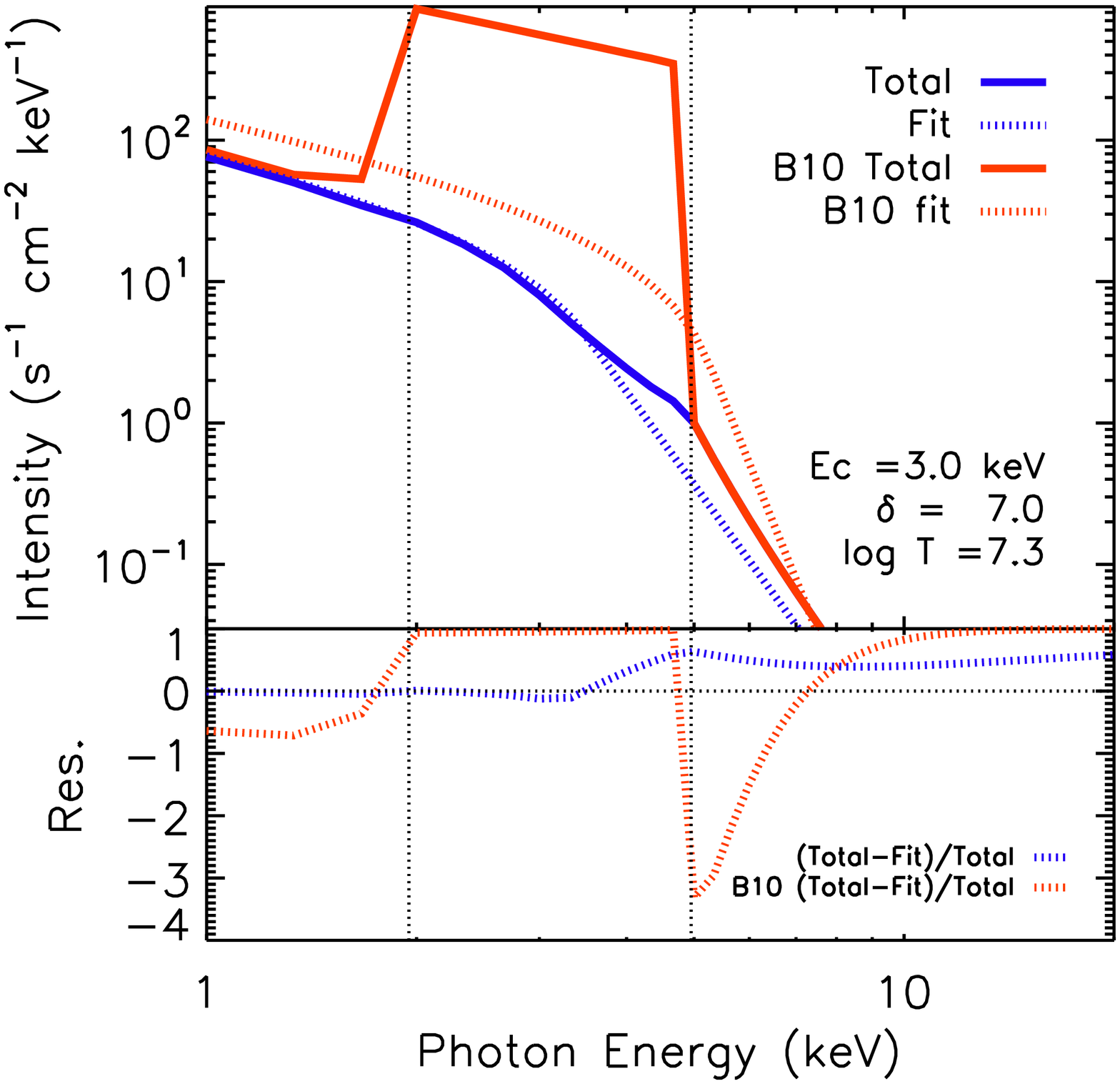}
\includegraphics[width=3.5in]{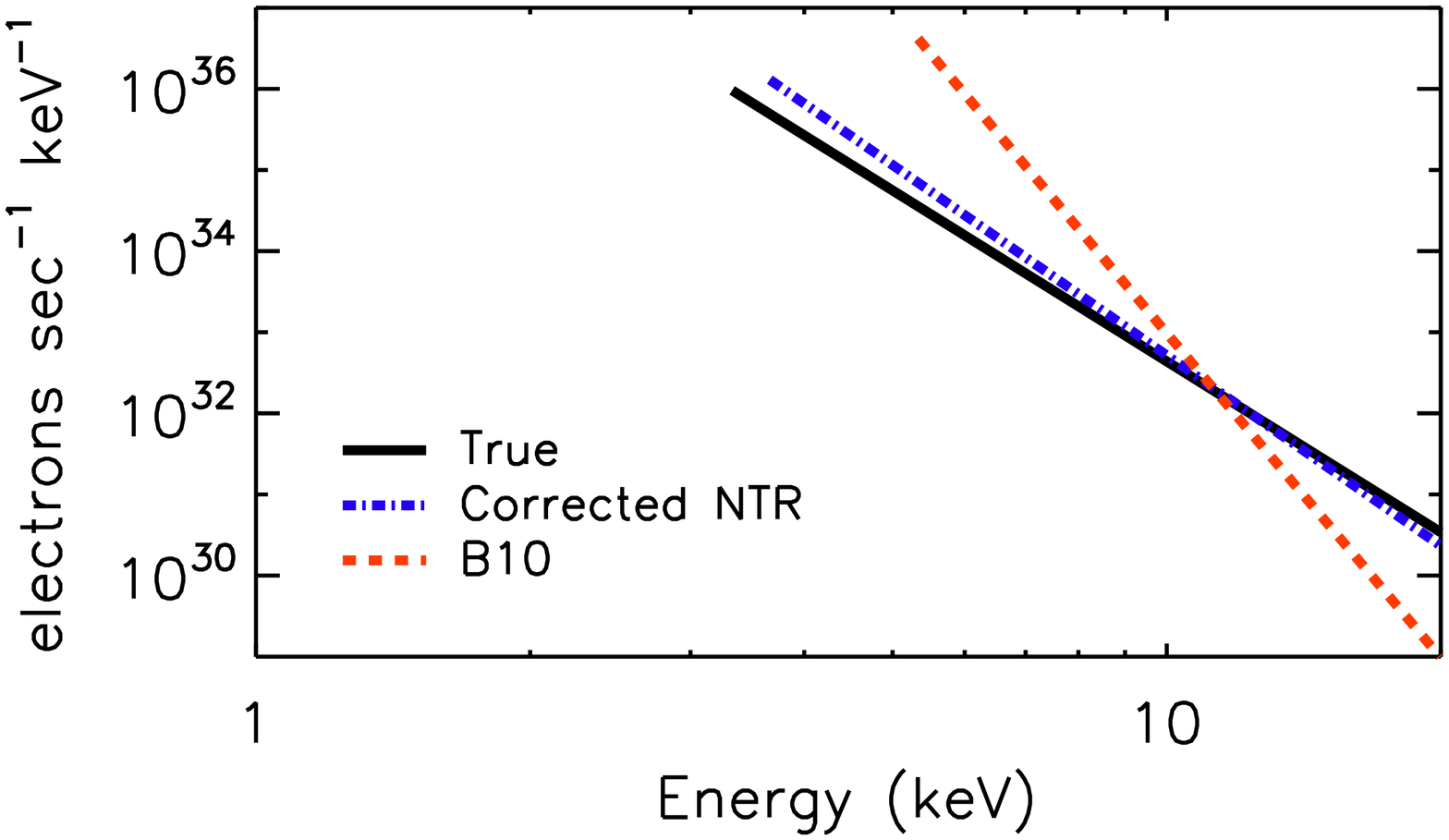}
\caption{Fitted photon spectra (top) with corrected NTR (blue) and the expression in B10 (red), with normalized residuals, defined as the total emission minus the fit emission, then divided by the total (center), and the derived electron distributions (bottom) found by fitting the photon spectra.  The solid line indicates the true electron distribution used to synthesize the photon spectra, with $E_{0c} = 3$\,keV, $\delta = 7$, and an electron flux above the cut-off $\mathfrak{F}_{0c} = 10^{36}$\,electrons sec$^{-1}$, while the other lines indicate the distributions found for each of the two cases.}
\label{fits}
\end{figure}

Figure \ref{fits} shows the photon spectra and their fits (top), the normalized residuals in the two cases, defined as the total emission minus the fit, then divided by the total emission (center), along with the derived electron distributions obtained from the fits, as compared to the true one used to synthesize the photon spectra (bottom).  The blue dotted line shows the distribution obtained where NTR was calculated with the corrected expression, with the following fit values obtained: $E_{0c} = 3.48$\,keV, $\delta = 7.79$, and $\mathfrak{F}_{0c} = 1.0 \times 10^{36}$ electrons sec$^{-1}$ (n.b. above the fitted cut-off 3.48\,keV), with a reduced chi-squared of 2.33.  The red dashed line similarly shows the case where NTR was calculated with the expression of B10, with the values $E_{0c} = 5.26$\,keV, $\delta = 13.3$, and $\mathfrak{F}_{0c} = 2.1 \times 10^{36}$ electrons sec$^{-1}$ (the fit was not good, with a reduced chi-squared $\approx 180$).  

In the case of the corrected NTR expression, the fit is reasonably good but shows physically important differences from the input, namely that $\delta$ and $E_{0c}$ are higher than the true value (7.8 compared to 7, and 3.5 versus 3.0\,keV), because of the extra contribution from NTR.  As a result, the emission at energies above 4\,keV are under-estimated.  In the B10 case, one important discrepancy is that the electron distribution obtained has significantly more electrons at energies $E_{0c,\text{fit}} < E < 10$\,keV than are actually present, due to the excess in recombination emission.  In both cases, the fitted electron rates $\mathfrak{F}_{0c}$ have a large discrepancy since they refer to different $E_{0c}$ and $\delta$ values.  At $3.6$\,keV, the corrected NTR case over-estimates the true injected electron spectrum by a factor of 2.6, while at $5.3$\,keV, the B10 case over-estimates the true spectrum by a factor of 116.  

\section{Interpretations}
\label{sec:interpretations}

There are several major implications of these results.  

\begin{enumerate}
\item \textbf{Spectra of microflares and small flares}  
In small events, electron acceleration is rather inefficient with small $E_{0c}$ and/or large $\delta$ ({\it e.g.} Figures 17 and 19 of \citealt{hannah2011}), with median values of $E_{0c} = 12$\,keV and $\delta = 8$ \citep{hannah2008b}.  Under such conditions, depending on temperature, NTR can be a significant contributor to the X-ray spectrum at small photon energies, dominating the NTB contribution for large enough $\delta$.  In such events therefore, NTR cannot be neglected in the fitting of low energy photon spectra (few to 10 keV or so).  
In these low energy bands, the fitting and interpretation of spectra are further complicated by the contributions not only of continuum bremsstrahlung and recombination from the thermal electrons but also of spectral lines from high $Z$ ions.  In large flares, with small $\delta$ and electrons accelerated to tremendous energies, bremsstrahlung is the dominant component of emission and NTR a relatively small consideration in fitting spectra.
\item \textbf{Inversion of hard X-ray data.}  
In larger flares, photon count statistics may be good enough to enable inversion of bremsstrahlung continuum to derive electron spectra \citep{brown1971}. If this is undertaken ignoring the small NTR contribution, the resulting errors in the electron spectrum can be large, especially when applied in the few to deka-keV range where the NTR contribution has discontinuous recombination spectral edges - cf. B08 Section 5. This issue is even more critical in the thick-target case addressed here where inference of $\mathfrak{F}_{0}(E_{0})$ from $J(\epsilon)$ involves not just the first but also the second derivative of  $J(\epsilon)$ \citep{brown1971}.
\item \textbf{Correction of Hard X-ray spectral fitting algorithms.}  
Existing Hard X-ray spectral fitting algorithms (e.g. RHESSI) incorporating the B08 and Corrigendum B10 for NTR contributions should be further amended with our results above.  
These amendments mostly reduce the importance of the NTR contribution compared with NTB but the ratio is extremely sensitive to model parameters and can be of order unity in some cases. We urge authors to use the scaling law in Section \ref{sec:results} to estimate the relative contribution of NTR to the non-thermal spectrum.  As explained in Section 5 of B08, recombination edges cause jumps in the derivative of the photon spectrum, which in turn cause errors in the inversion of the non-thermal electron distribution, even when the contribution of NTR to the total spectrum is relatively small.  Consequently, we re-emphasize the conclusion of B08 that it is important to include NTR in spectral data fitting and interpretations.
\end{enumerate}

There are a few facets of the non-thermal spectrum that remain to be considered.  Importantly, we have not treated the non-thermal excitation of bound-bound transitions, or the downward cascade of electrons after recombining into a level with $n > 1$.  We will evaluate the relative importance of these mechanisms in future work, in order to fully understand the non-thermal spectrum in flares. \\

\vspace{5mm}
\noindent {\it Acknowledgements} This research was performed while JWR held an NRC Research Associateship award at the US Naval Research Laboratory with support from NASA.  JCB is grateful to JWR for picking up on the further error in the Erratum \citep{brown2010} and acknowledges the support of an STFC Consolidated Grant.  We thank Harry Warren for suggestions that improved this work, particularly in regards to efficient fitting methods.  We also thank the anonymous referees for comments that clarified and improved the content of this paper.  CHIANTI is a collaborative project involving George Mason University, the University of Michigan (USA) and the University of Cambridge (UK).


\begin{thebibliography}{}
\bibitem[Bethe \& Heitler(1934)]{bethe1934} Bethe, H., \& Heitler, W. 1934, RSPSA, 146, 83
\bibitem[Brown(1971)]{brown1971} Brown, J.C.\ 1971, \solphys, 18, 489
\bibitem[Brown \& Mallik(2008)]{brown2008} Brown, J.C., \& Mallik, P.C.V.\ 2008, \aap, 481, 507
\bibitem[Brown \& Mallik(2009)]{brown2009} Brown, J.C., \& Mallik, P.C.V.\ 2009, \apjl, 697, L6
\bibitem[Brown, Mallik \& Badnell(2010)]{brown2010} Brown, J.C., Mallik, P.C.V., \& Badnell, N.R.\ 2010, \aap, 515, C1
\bibitem[Del Zanna et al.(2015)]{delzanna2015} Del Zanna, G., Dere, K.P., Young, P.R., et al.\ 2015, \aap, 582, 56
\bibitem[Dere et al.(1997)]{dere1997} Dere, K.P., Landi, E., Mason, H.E., et al.\ 1997, \aaps, 125, 149
\bibitem[Fleishman et al.(2011)]{fleishman2011} Fleishman, G.D., Kontar, E.P., Nita, G.M., \& Gary, D.E.\ 2011, \apjl, 731, L19
\bibitem[Fleishman et al.(2016)]{fleishman2016} Fleishman, G.D., Pal'shin, V.D., Meshalkina, N., et al.\ 2016, \apj, in press.
\bibitem[Grigis \& Benz(2004)]{grigis2004} Grigis, P.C., \& Benz, A.O.\ 2004, \aap, 426, 1093
\bibitem[Hannah et al.(2008a)]{hannah2008} Hannah, I.G., Krucker, S., Hudson, H.S., et al.\ 2008, \aap, 481, L45
\bibitem[Hannah et al.(2008b)]{hannah2008b} Hannah, I.G., Christe, S., Krucker, S., et al.\ 2008, \apj, 677, 704
\bibitem[Hannah et al.(2011)]{hannah2011} Hannah, I.G., Hudson, H.S., Battaglia, M., et al.\ 2011, \ssr, 159, 263
\bibitem[Koch \& Motz(1959)]{koch1959} Koch, H.W., \& Motz, J.W., 1959, RvMP, 31, 920
\bibitem[Kramers(1923)]{kramers1923} Kramers, H.A.\ 1923, Phil. Mag., 46, 836
\bibitem[Markwardt(2009)]{markwardt2009} Markwardt, C.~B.\ 2009, Astronomical Data Analysis Software and Systems XVIII, 411, 251
\bibitem[Masuda et al.(2013)]{masuda2013} Masuda, S., Shimojo, M., Kawate, T., et al.\ 2013, PASJ, 65, 1
\bibitem[O'Flannagain et al.(2013)]{oflannagain2013} O'Flannagain, A.M., Gallagher, P.T., Brown, J.C., et al.\ 2013, \aap, 555, 21
\end{thebibliography}
\end{document}